\pdfoutput=1
\documentclass[epj]{webofc}
\usepackage[varg]{txfonts}   
\wocname{EPJ Web of Conferences}
\woctitle{ICNFP 2015}
\begin{document}
\selectlanguage{english}
\title{Thermalization and confinement in strongly coupled gauge theories}
%
%

\author{Takaaki Ishii\inst{1,2} \and
        Elias Kiritsis\inst{1,3} \and
        Christopher Rosen\inst{1,4}
}

\institute{Crete Center for Theoretical Physics, University of Crete, Heraklion 71003, Greece
\and
University of Colorado, Boulder, CO 80309, USA	
\and
University of Paris Diderot, Sorbonne Paris Cit\'{e}, APC, Paris F-75205, France
\and
Imperial College, London SW7 2AZ, UK
}

\abstract{%
Quantum field theories of strongly interacting matter sometimes have a useful holographic description in terms of the variables of a gravitational theory in higher dimensions. This duality maps time dependent physics in the gauge theory to time dependent solutions of the Einstein equations in the gravity theory. In order to better understand the process by which ``real world'' theories such as QCD behave out of thermodynamic equilibrium, we study time dependent perturbations to states in a model of a confining, strongly coupled gauge theory via holography. Operationally, this involves solving a set of non-linear Einstein equations supplemented with specific time dependent boundary conditions. The resulting solutions allow one to comment on the timescale by which the perturbed states thermalize, as well as to quantify the properties of the final state as a function of the perturbation parameters. We comment on the influence of the dual gauge theory's confinement scale on these results, as well as the appearance of a previously anticipated universal scaling regime in the ``abrupt quench'' limit.
}

\begin{flushright}
CCQCN-2016-118\\
CCTP-2016-1\\
Imperial/TP/2015/CR/01\\
\vspace*{-0.8cm}
\end{flushright}

\maketitle

\section{Introduction}
\label{sec:intro}
\begin{figure}
\centering
     \includegraphics[width=.6\textwidth]{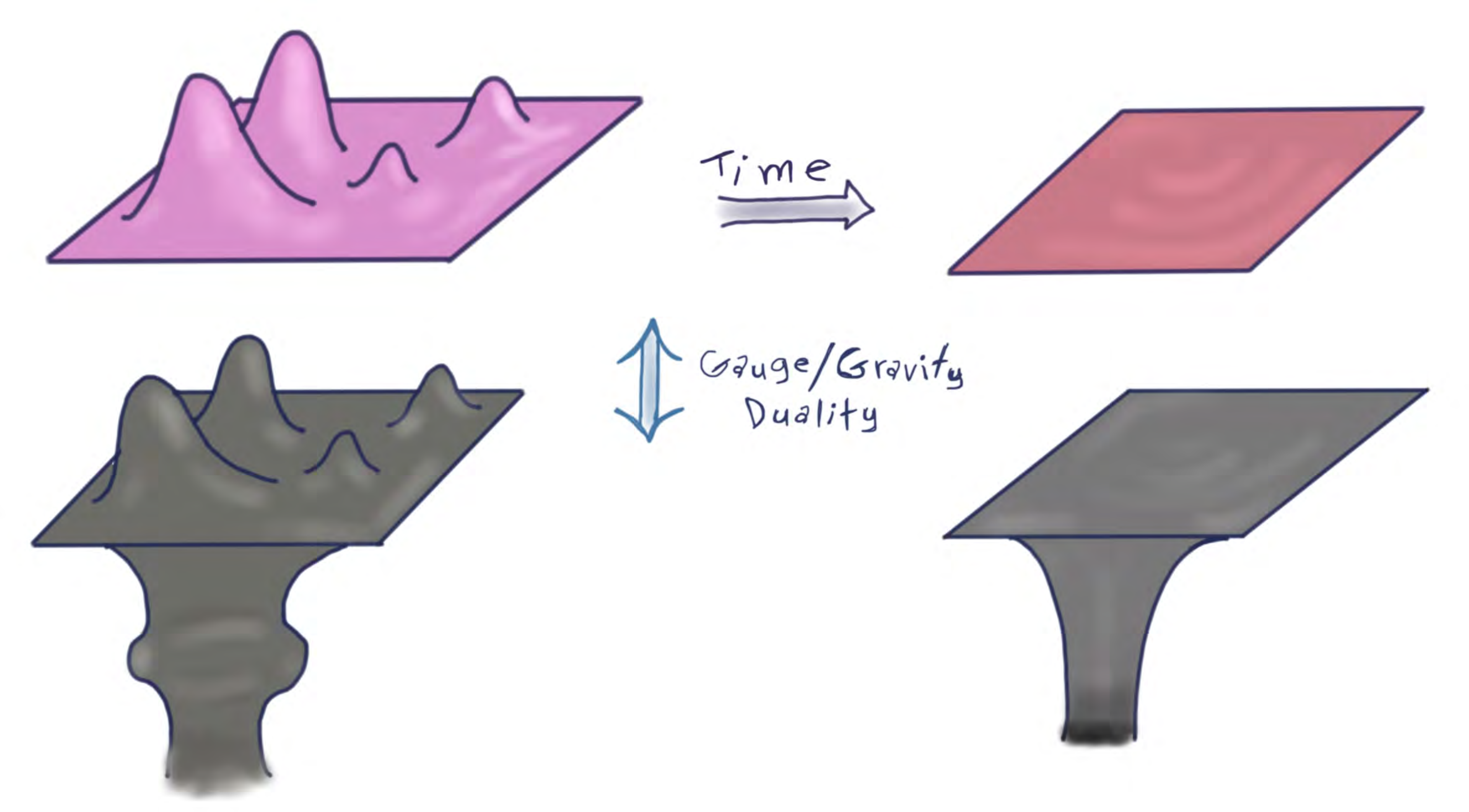}
     \caption{Cartoon depiction of holographically dual thermalization processes. On the top, a strongly coupled field theory is driven out of equilibrium by an external probe, and at late times may attain thermal equilibrium. On the bottom, the dual process is understood as a perturbation of a gravitational theory in one higher dimension, and late time thermal equilibration corresponds to black hole formation in the bulk.}
     \label{fig:thermDual}
\end{figure}

The dynamical properties of strongly interacting matter are interesting both because relatively little is currently known about them, as well as the fact that they are increasingly accessible in the laboratory and large-scale experiments like the heavy ion collisions at RHIC and LHC. One standard method for calculating in strongly coupled field theories is to regularize the theory on a lattice. This method however relies on a Euclidean formulation of the field theory, and hence it is rather difficult to make contact with real-time dynamics. Finding a way around this obstacle takes special importance in the context of ongoing programs in the relativistic heavy ion collisions. At the energy scales currently accessible, the hot, dense matter produced in such collisions of heavy nuclei behaves as a strongly interacting plasma and thus novel theoretical approaches are needed to characterize its dynamical properties. One such approach is based on a holographic duality between gauge theories and string theories illustrated in figure \ref{fig:thermDual}. This approach was employed in \cite{Ishii:2015gia} to model the dynamical response of a confining gauge theory at strong coupling, and that work provides the foundation for these proceedings.

\section{Thermalization in Holography}
\label{sec:therm}
Before developing the holographic model in any detail, it is useful to orient oneself with some very general expectations from the field theory perspective. Given any strongly coupled medium, a useful characterization of its properties is provided by its response to a variety of external perturbations. If the dynamics of the medium can be captured by a quantum field theory effectively described by the Lagrangian $\mathcal{L}_{\mathrm{QFT}}$, then the application of a perturbation by a local operator $\mathcal{O}$ in a manner described by the source $f_0$ is accomplished by adding to the Lagrangian
\begin{equation}
\mathcal{L}_{\mathrm{QFT}}\to \mathcal{L}_{\mathrm{QFT}}+f_0\mathcal{O}.
\end{equation}
In the special case that the source is isotropic in space and changes only in time, the probe will do work on the system as dictated by the following Ward identity:
\begin{equation}\label{eq:ward}
\nabla^t\langle T_{tt}\rangle = \dot{f}_0 \langle\mathcal{O}\rangle.
\end{equation}
Here the ``dot'' denotes a partial derivative with respect to time and $T$ is the stress-energy tensor.

The response of the system to the time dependent perturbation can be measured in any of a number of different ways, and these different methods generically elucidate different aspects of the system's response. A particularly natural measure is  provided by the various one point functions of gauge invariant operators of the field theory. By examining the dependence of these expectation values on time, one can access both the endpoint of the dynamical evolution, as well as the route the system takes to arrive there.

One scenario that will be of particular importance in the present work arises when the late time behavior of the system is independent of time, and the various one point functions take the form
\begin{equation}
\langle\mathcal{O}(t\to\infty)\rangle \sim \mathrm{tr}\,\rho\mathcal{O},
\end{equation}
where $\rho$ is a thermal distribution. If this is the case, then the endpoint of the evolution is a thermal state, and the equilibration process is called ``thermalization''.

For strongly coupled field theories with a holographic dual the thermalization process can alternatively be described in terms of weakly coupled geometric variables. These are the bulk fields of a gravitational theory in one higher dimension. In terms of these variables, thermalization in a 3+1 dimensional field theory is understood from the dual viewpoint as horizon formation in a 4+1 dimensional theory with gravity.

\begin{figure}[t]
\centering
     \includegraphics[width=.5\textwidth]{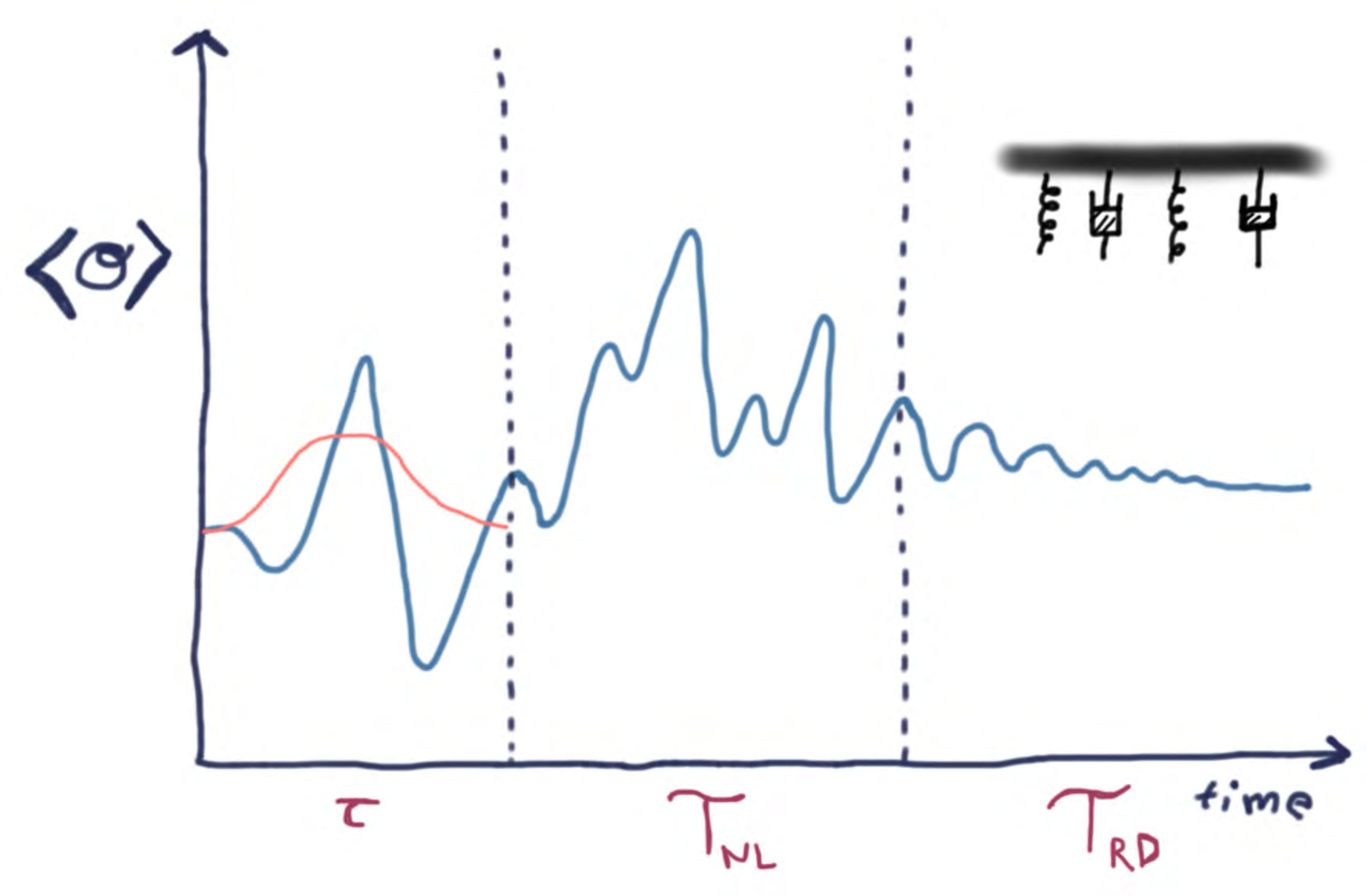}
     \caption{Typical response of a holographic system to a time-dependent perturbation. The vertical axis is the expectation value of the scalar operator determined by the behavior of the scalar  near the boundary, while time is along the horizontal axis. After a time $\tau$ the system is driven out of equilibrium, and may pass through any number of non-linear regimes $\mathcal{T}_\mathrm{NL}$ before ``ringing down'' to thermal equilibrium with a characteristic time scale $\mathcal{T}_\mathrm{RD}$. The ring-down regime is anticipated from the well-known fact that black branes respond to small (linearized) perturbations much like a system of damped oscillators (inset). }
     \label{fig:bbRes}
\end{figure}

The formulation of the dynamical process in terms of the dual gravity theory provides advantages both in terms of computation and intuition. In figure \ref{fig:bbRes}, a generic expectation for the dynamical evolution of a holographic one point function is sketched. To characterize this evolution, it is useful to imagine dividing the response into different regimes each controlled by any of an assortment of different ``characteristic'' time scales. For instance, if one imagines varying the source $f_0$ over a timescale $\tau$, then $\tau$ is roughly the characteristic time scale governing the amount of time taken to drive the system out of equilibrium. The subsequent non-equilibrium system will evolve according to the non-linear gravitational equations of motion, and thus it could be possible that the system will pass through a highly non-linear regime, $\mathcal{T}_\mathrm{NL}$. This time scale could conceivably be further subdivided into any number of distinct regimes.

If a horizon forms in the bulk, it is natural to expect that a ``ring-down'' regime $\mathcal{T}_\mathrm{RD}$ defines the system's near-equilibrium behavior for the dual theory to thermalize at late times. This expectation is based on the fact that the linearized response of the final state black brane to small perturbations is controlled by the lowest lying (most real) quasi-normal mode in the spectrum at late times.

As a consequence of the preceding discussion, it is sensible to define a thermalization time as simply the sum of all the characteristic time scales defining the dynamic evolution except the time scale characterizing the application of the external source,
\begin{equation}
\mathcal{T}_\mathrm{THERM} = \mathcal{T}_\mathrm{NL}+\ldots +  \mathcal{T}_\mathrm{RD}.
\end{equation}
Thus defined, the thermalization time depends only on the out-of-equilibrium response and relaxation of the strongly coupled matter. In the majority of holographic examples of thermalization in a strongly coupled field theory found in the literature, the thermalization time appears to be dominated by the ring-down regime, i.e $\mathcal{T}_\mathrm{THERM}\approx\mathcal{T}_\mathrm{RD}$.

\section{Confinement in Holography}
\label{sec:confine}
One feature absent from many previous holographic investigations of thermalization is the presence of a confinement scale analogous to the $\Lambda_\mathrm{QCD}$ of quantum chromodynamics. It is natural to wonder if the existence of such an additional  characteristic energy scale could alter the dynamical response of the theory in important ways, perhaps leading to thermalization processes in the dual field theory that lie outside the omnipresent ring-down regime.

The introduction of confinement to holographic gauge theories has a long history \cite{Witten:1998zw,Kinar:1998vq}, and is by now a familiar aspect of gauge/gravity duality. Confinement in any gauge theory can be identified with an ``area law'' behavior of the Wilson loop $\mathcal{W}$,
\begin{equation}
\langle\mathcal{W} \rangle \sim e^{-F_s \cdot T L}
\end{equation}
where $F_s$ is the QCD string tension, and $TL$ is the area of the Wilson loop describing the separation by length $L$ of a quark anti-quark pair propagating for time $T$. In a holographic theory, the Wilson loop has a dual description in terms of a certain minimal surface drooping into the bulk geometry. It can be shown \cite{Kinar:1998vq} that to obtain the desired area law behavior of the boundary theory Wilson loop, the bulk geometry must obey certain requirements.  For example, if the bulk string frame metric is written as
\begin{equation}
\mathrm{d}s^2 = e^{2A(z)}\left(-\mathrm{d}t^2+\mathrm{d}\vec{x}\,^2 + \mathrm{d}z^2\right)
\end{equation}
where $z$ is the radial coordinate orthogonal to the boundary theory directions, then the dual Wilson loop will only exhibit area law behavior if the warp factor $A(z)$ achieves a minimum at some radial coordinate $z_0$, and if $A(z_0)$ is finite.

\begin{figure}
\centering
     \includegraphics[scale=0.22]{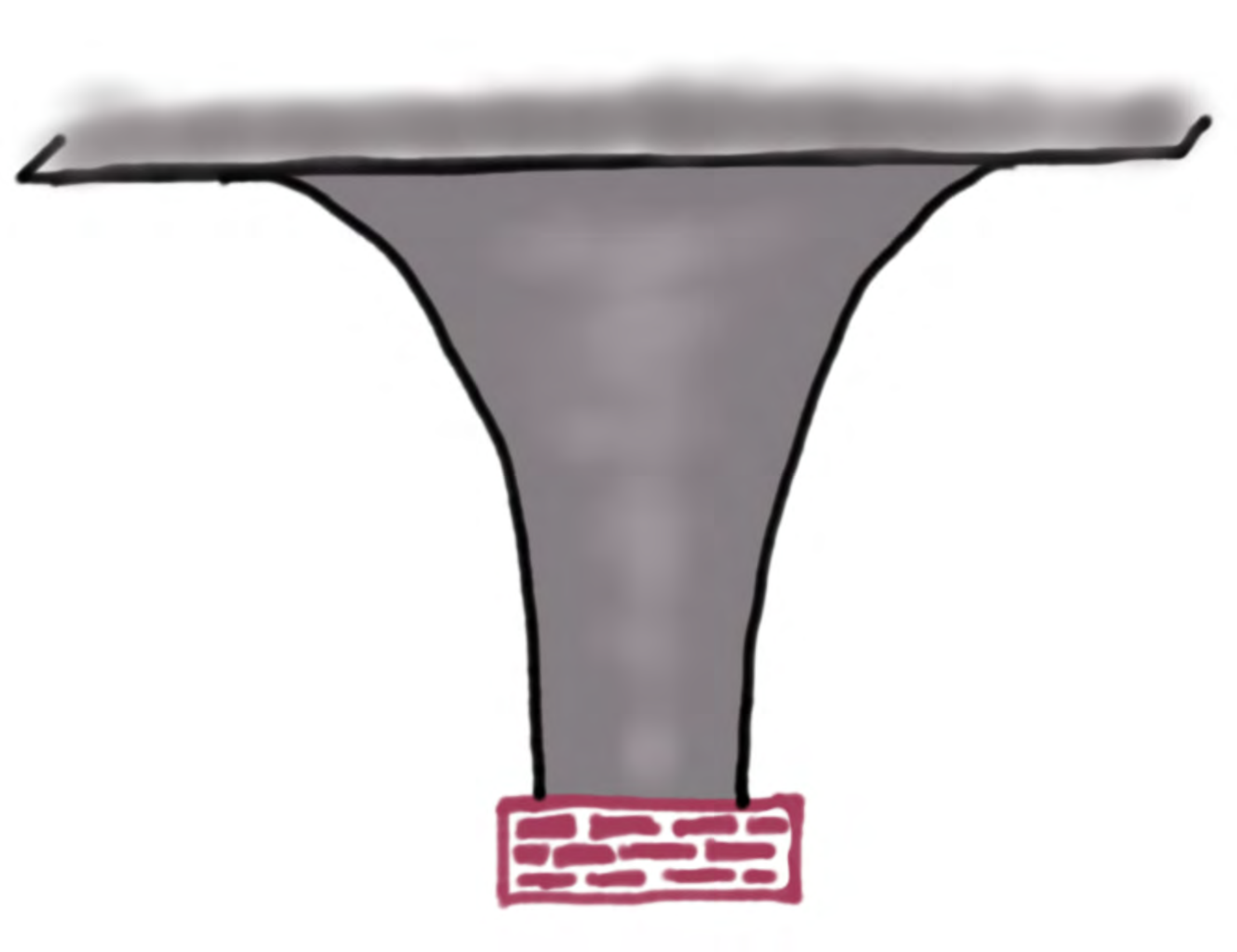}\hspace{1cm}
     \includegraphics[scale=0.22]{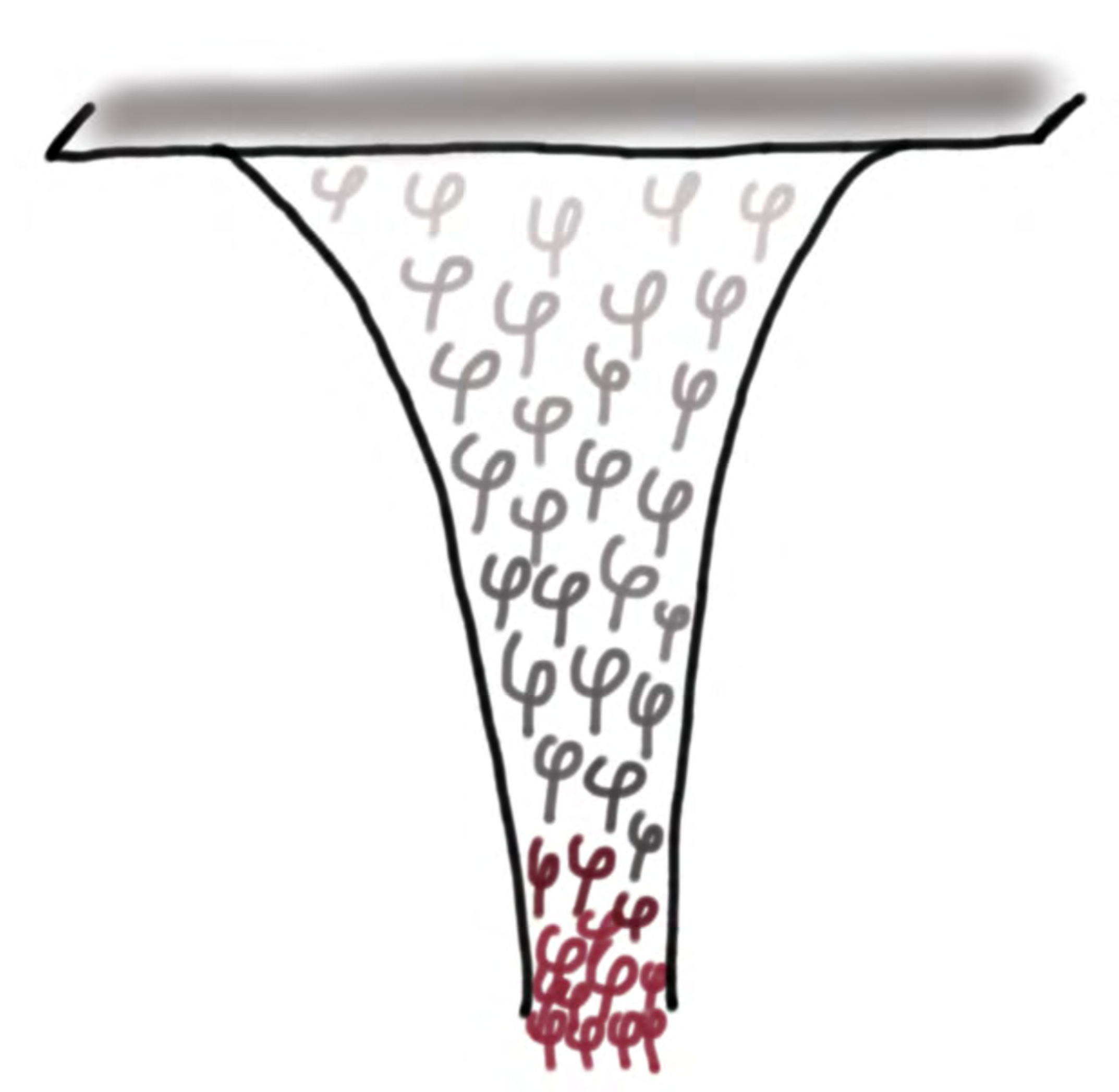}
     \caption{Examples of gravitational systems holographically dual to confining gauge theories. In the hard wall model (left) the geometry is artificially cut off at some radial position $z = z_0$, here shown as a brick wall. In bulk theories with scalar matter like the Einstein-Dilaton theories (right), it is possible to engineer a potential for the scalar which effectively shields probes from reaching arbitrarily far into the throat.}
     \label{fig:capG}
\end{figure}

Thus, roughly speaking, gravitational theories dual to confining gauge theories are ``capped'' in the radial direction. The mechanism for capping the geometry can be either abrupt, as in the case of the so-called hardwall model , or gentle perhaps with the aid of other bulk fields. These possibilities are illustrated and explained in more detail in figure \ref{fig:capG}.

Holographic thermalization in the hard-wall model was explored initially in \cite{Craps:2013iaa, Craps:2014eba}. The primary result was that the introduction of a confinement scale indeed leads to a more complex thermalization processes. More specifically, a class of perturbations were identified that {\it never} thermalize \cite{Craps:2013iaa}. This can be understood as a consequence of the following simple argument: if the perturbation to the hard-wall geometry carries sufficient energy density to produce a black brane with horizon above (closer to the boundary than) the harwall at $z_0$, then the perturbation is indifferent to the existence of the wall and will thermalize in a time $\mathcal{T}_\mathrm{THERM} \approx \mathcal{T}_\mathrm{RD}$. If the associated energy density is lower, the perturbation will scatter from the hard-wall and return to the boundary where it scatters once more. This process appears to repeat for as long as the numerical routines of \cite{Craps:2014eba} were able to run. Recently, analogous calculations were performed in the AdS soliton background \cite{Craps:2015upq} which is qualitatively similar to the hardwall in that there is a mass gap for black brane formation in the bulk.  There too perturbations were identified which appear to scatter indefinitely, while others lead to collapse and black-brane formation.

\section{Dynamical Quenches in a Confining Gauge Theory}
\label{sec:dynam}
Clearly, an important next step is to determine whether or not something similar can happen in more realistic holographic models of confining gauge theories. An example of one such class of models is the Einstein-Dilaton class, whose solutions minimize an action of the form
\begin{equation}\label{eq:act}
S = \frac{1}{2\kappa^2}\int\mathrm{d}^5 x\sqrt{-g}\left( R-\frac{4}{3}(\partial\varphi)^2+V(\varphi)\right)-\frac{1}{\kappa^2}\int_\partial\mathrm{d}^4x\sqrt{-\gamma}\mathcal{K} + S_{\mathrm{CT}}.
\end{equation}
The first term in this action describes the interactions of a scalar field $\varphi$ coupled to gravity, subject to the scalar potential $V$, while the second term is the familiar Gibbons-Hawking-York boundary term. The final term  denotes a collection of bulk counterterms that must be added to the action to regularize the action on-shell, and allows for the computation of holographically renormalized correlation functions.

In a ``top down'' holographic model,   the scalar potential $V$ would be fully determined by the symmetries of an underlying supergravity theory. Top down models often have the advantage of providing a gravitational theory whose bulk fields are dual to explicitly identifiable operators in  a known boundary gauge theory. Constructing such models with the desired dual physics is typically hard work, however, and  it is often considerably more pragmatic to  employ a ``bottom up'' approach.  In a bottom up model, the scalar potential can be arbitrarily tuned to induce the desired dual physics. The freedom to tune this potential comes at the cost of rigor, as one subsequently forfeits the ability to make detailed claims about the name and operator content of the dual field theory. Nonetheless, in the present case, we will adopt a bottom up perspective and choose $V$ such that the dual boundary gauge theory is confining at low temperatures and such that solutions with non-zero scalar correspond to relevant deformations by a dimension three scalar operator (for details, see \cite{Ishii:2015gia}). An example of a scalar potential that realizes these features is given by
\begin{equation}\label{eq:pot}
V(\varphi)=\frac{12(1+a\varphi^2)^{1/4}\cosh\frac{4}{3}\varphi-b\varphi^2}{L^2}
\end{equation}
with the parameter choice $(a,b) =(1/500, 10009/1500)$ and $L$ the AdS scale.

The solutions to this bulk theory describe the various phases available to the dual field theory. At zero temperature, the dual gravitational solution is a horizon-less geometry with a running scalar  and a fairly innocuous naked singularity in the bottom of the throat. Ideally, one would like to perturb this geometry by varying the source for the dual scalar operator in time. This would roughly correspond to driving the system out of its ground state by altering the coupling of a dimension three scalar operator. In practice, the rapidly diverging bulk scalar greatly complicates the numerical analysis and it is necessary to introduce an IR cutoff deep in the throat region of the geometry.

\begin{figure}
\centering
\includegraphics[scale=0.6]{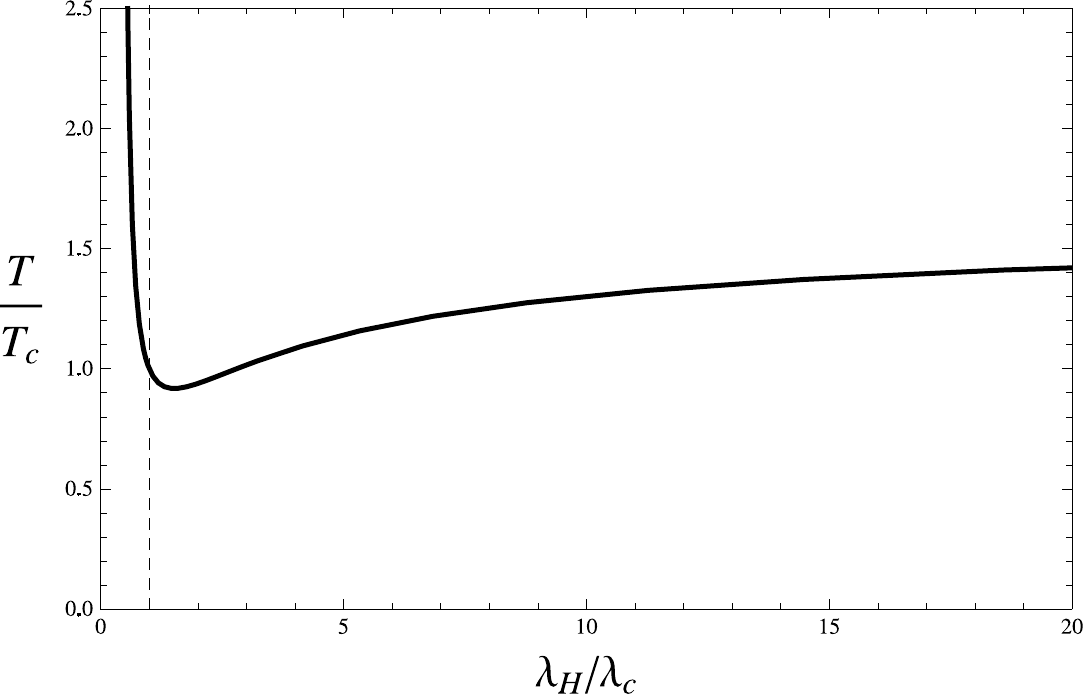}
\caption{Plot of the temperature of the black hole solution scaled by the critical temperature as a function of the value of the scalar at the horizon, $\lambda_H/\lambda_c$.  At $\lambda_H = \lambda_c$ there is the first order phase transition between the large black hole and the thermal gas solution which does not have a horizon, while the small black hole co-exists as a thermodynamically disfavored solution with a small horizon.}
\label{fig:Told3}
\end{figure}

A sensible means for introducing such a cutoff can be found in another branch of solutions to the bulk action.  A particularly well suited branch contains the ``small black hole'' solutions, which shield the singularity behind a very small horizon. These solutions provide initial state candidates in which all bulk fields are well behaved between the UV boundary and the horizon in the IR. In this way the small horizon regulates the IR, but not without introducing some drawbacks.  First, the introduction of a horizon in the initial state obviously renders any  information related to horizon formation beyond the reach of our computations. Second, the small black holes along this branch are never the thermodynamically preferred solutions in the dual field theory. Nonetheless, they have the advantage of retaining much of the flavor of the zero temperature solution (notably a scalar which grows rapidly in the IR) while shielding the numerical methods from the existence of the IR singularity. In figure~\ref{fig:Told3} the finite temperature solutions to the model (\ref{eq:act}) are shown, parametrized by the value of the scalar $\lambda = e^\varphi$ at the horizon. The small black holes are the branch to the right of the minimum temperature solution.

Our computational strategy will be as follows: beginning from a small black hole initial state, we turn on a time dependent source for the boundary scalar operator of the form
\begin{equation}\label{eq:f0fn}
f_0(t) = \tilde{f}_0 \left(1 + \tilde{\delta}\,e^{-\frac{v^2}{2\tilde{\tau}^2}}\right),
\end{equation}
which can be tuned by varying $\tilde{\delta}$ and $\tilde{\tau}$. This change in the scalar at the boundary propagates into the bulk, backreacting on the geometry. This dynamical process is fully described by the non-linear Einstein equations, whose causal time evolution must be solved numerically step by step in time until a final steady state solution has been obtained. At each boundary time, a standard application of the holographic dictionary allows one to read off the one-point functions of the stress-energy tensor and scalar operator in the dual gauge theory.

Evolving the solutions using the Einstein equations is difficult and requires an assortment of numerical methods tailored to the gravitational problem. While various numerical schemes have found success in the gravitational literature, we have found the characteristic formulation particularly convenient. This formulation has previously been used with success in a variety of holographic computations, starting with the important work of \cite{Chesler:2008hg}. In this setup, the Einstein equations are recast as a nested collection of ordinary differential equations that can be solved in series when supplemented with some initial data.  Independent of the choice of scheme, one must also decide which numerical techniques are best suited to perform the actual integration of the equations of motion. One option, which has received much attention in the numerical holography literature, are the pseudo-spectral methods. Under certain circumstances, these methods offer excellent convergence properties and are comparatively inexpensive computationally. These benefits are, unfortunately, somewhat compromised in the case that  the bulk fields contain non-analytic behavior within the computational domain. Therefore, we found it best to utilize an assortment of finite difference schemes to integrate our solutions into the bulk and forward in time.

\begin{figure}
\centering
     \includegraphics[scale=0.7]{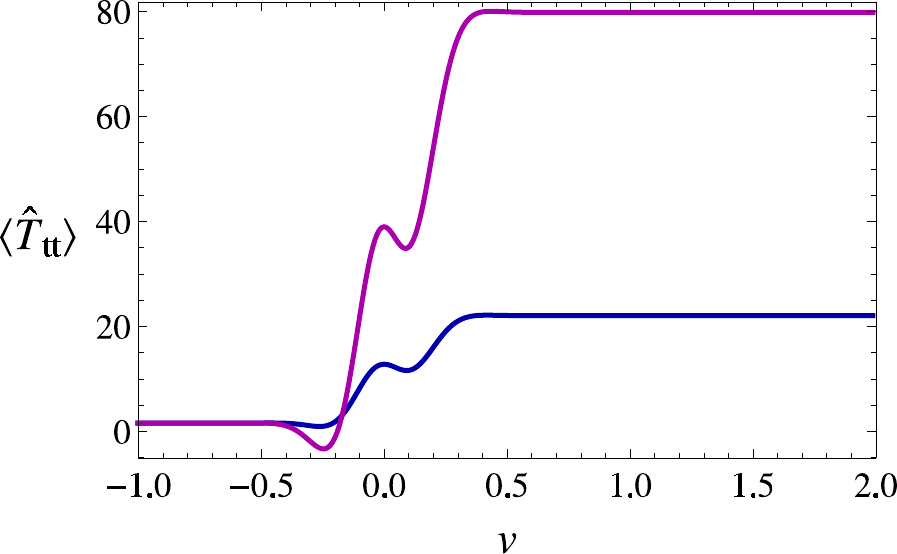}
     \includegraphics[scale=0.7]{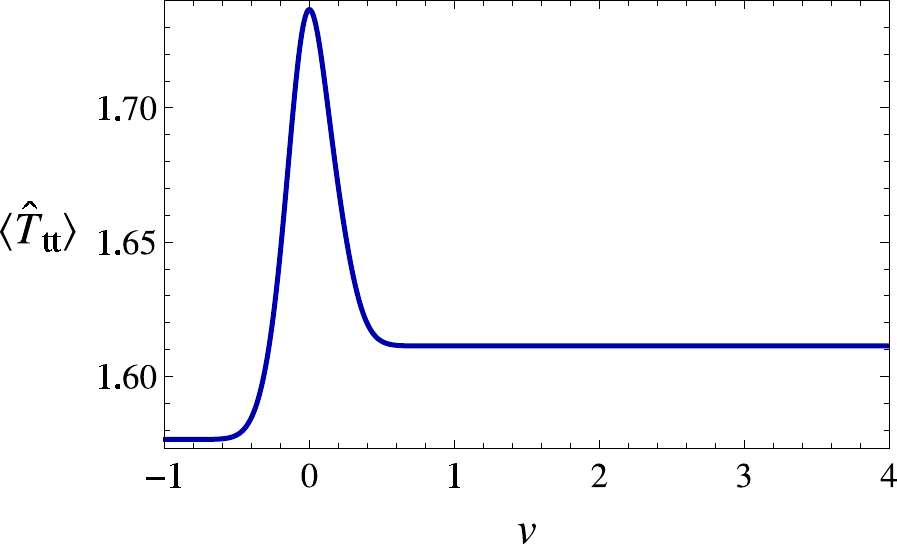}
     \caption{Time evolution of $\langle T_{tt}\rangle$ for large amplitude (left) and small amplitude (right) quenches in the small black hole initial state.}
     \label{fig:evo}
\end{figure}

Upon successful evolution of the Einstein equations, the system's response reveals an energy density which changes markedly on the time scale of the quench $\tilde{\tau}$, followed by a rapid equilibration to its final state value. Some examples of this evolution are plotted in figure \ref{fig:evo}. By construction, this rapid change satisfies the Ward identity \eqref{eq:ward} in the presence of the time dependent source of the form \eqref{eq:f0fn}. Indeed, both the Ward identity governing the divergence of the boundary stress energy tensor as well as the Ward identity relating the trace of this stress tensor to classical and anomalous conformal symmetry breaking terms are guaranteed to hold in any solution to the gravitational equations of motion.

\begin{figure}
\centering
     \includegraphics[scale=0.95]{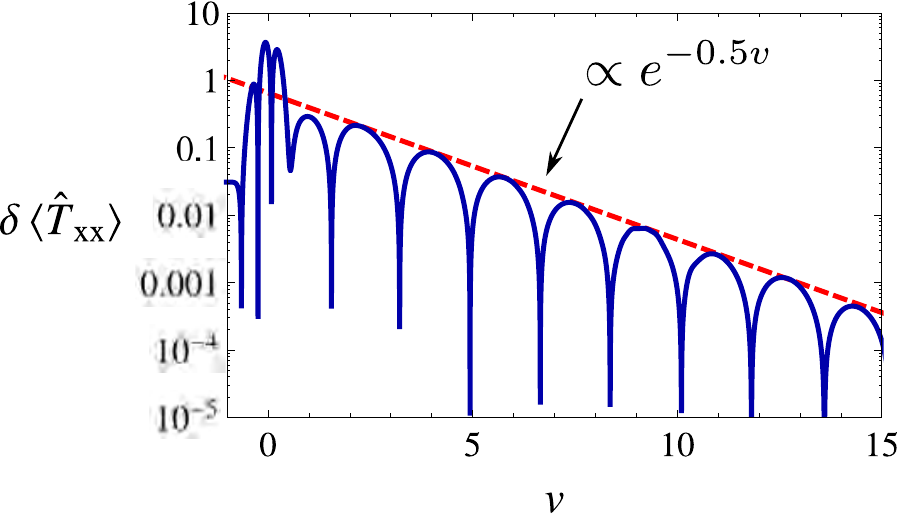}
     \caption{Generic late-time behavior of the magnitude of a one-point function's deviation from its equilibrium value. After the quench time $\sim \tilde{\tau}$ the response equilibrates much like a damped oscillator. The damped oscillations correspond to the excitations of the gravitational system's lowest lying quasi-normal mode.}
     \label{fig:osc}
\end{figure}

To  better understand the rapid approach of the perturbed system to thermal equilibrium, it is useful to look in more detail at the late time behavior of the system. In figure \ref{fig:osc} the time dependence of the scalar one point function is plotted on a logarithmic scale. One notes immediately that after a short time (again controlled by the quench width $\tilde{\tau}$), the system oscillates with a well defined frequency $\omega_{*}$ while simultaneously approaching its equilibrium value exponentially with decay constant $\Gamma$. This is exactly as one would expect from a linear system controlled by an excited mode of the form $\omega_1 = \omega_{*}-i\,\Gamma$. Evidently, even in the confining model of (\ref{eq:act}-\ref{eq:pot})  the thermalization time seems to be dominated by the linear regime, $\mathcal{T}_\mathrm{THERM}\approx \mathcal{T}_\mathrm{RD}$.

\begin{figure}[t]
\centering
\includegraphics[scale=0.8]{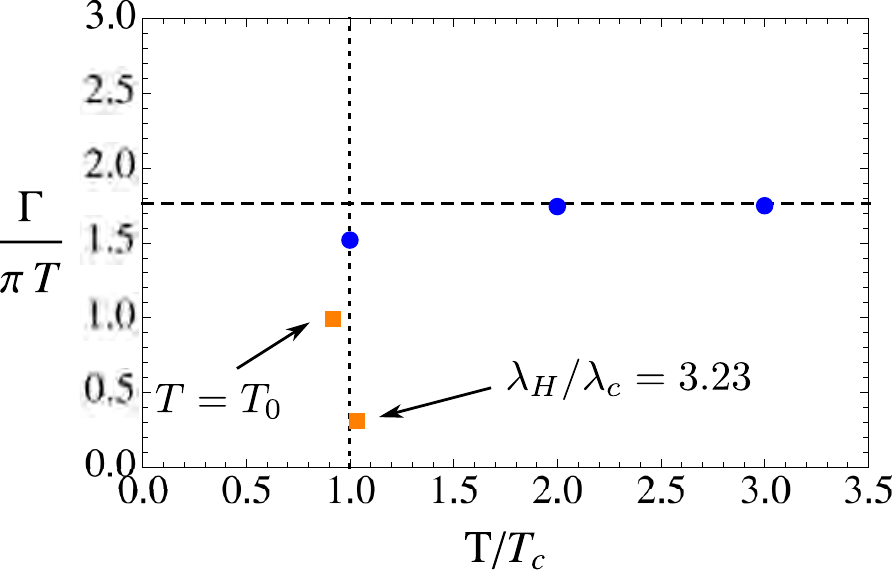}
\caption{The temperature dependence of the decay width $\Gamma$ for the lowest-lying scalar quasi-normal mode in several states of our theory. The blue circles are large black-branes whose temperature is an integer multiple of $T_c$. The orange squares correspond to the minimum temperature black brane at $T=T_0$ (top) and the smallest black hole we perturb in our study at $\lambda_H/\lambda_c=3.23$ (bottom). The ratio $\Gamma/\pi T$ approaches 1.75953 (the dashed line) at high temperatures, which coincides with the expected value for perturbations of AdS$_5$ Schwarzschild by a dimension 3 scalar operator \cite{Nunez:2003eq}. }
\label{fig:QNM}
\end{figure}

The decay width $\Gamma$ is given by the imaginary part of the lowest lying scalar quasi-normal mode of the system's final state black hole. In figure \ref{fig:QNM} we quantify this by plotting $\Gamma$ as a function of temperature for several thermal states of our holographic theory. As anticipated, linear scaling of $\Gamma$ with temperature appears in the conformal (small scalar) limit, and deviations from this behavior are already evident at the first order phase transition where $T=T_c$. These deviations become larger as one moves further onto  the small black hole branch. In this way $\Gamma$ quantitatively captures the degree of the conformal symmetry breaking in the presence of the nontrivial confining scalar potential \eqref{eq:pot}.

\begin{figure}
\centering
     \includegraphics[scale=0.7]{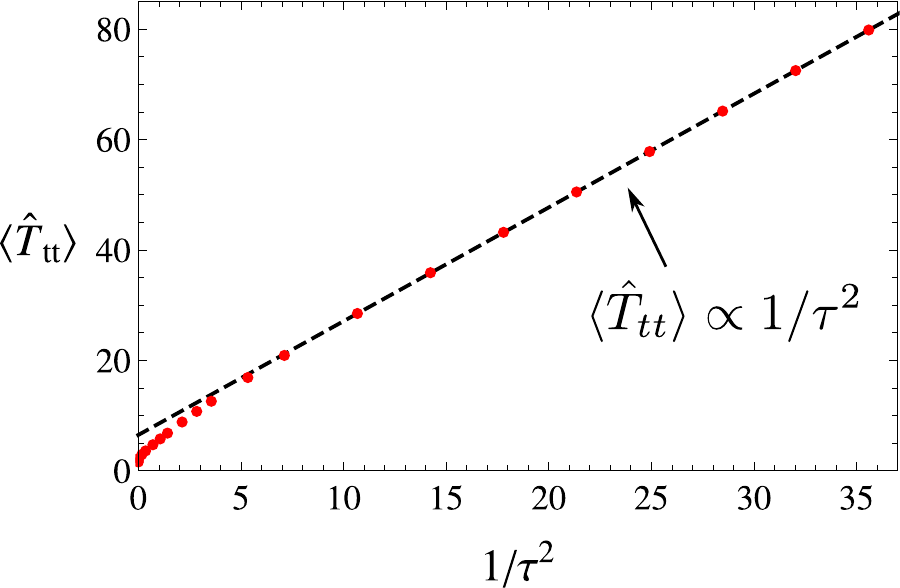}
     \includegraphics[scale=0.7]{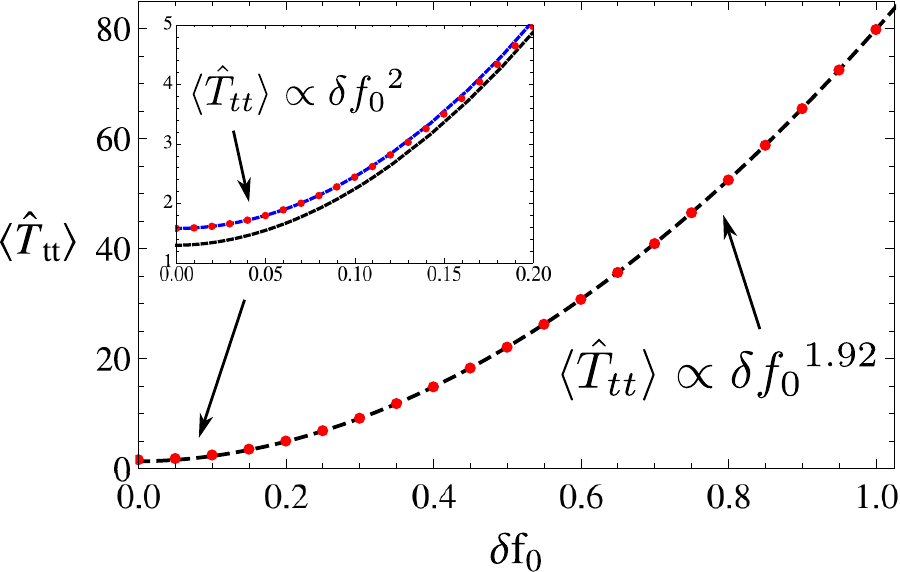}
     \caption{Final state energy density $\langle T_{tt}\rangle$ as a function of quench parameters. In the left plot the quench amplitude is held fixed, while in the right plot the quench duration is fixed.}
     \label{fig:Tfvp}
\end{figure}

Outside of thermalization times, the holographic approach can also provide some insight into the dependence of the final state on parameters of the perturbation. In figure \ref{fig:Tfvp} the final state energy density is plotted as a function of quench duration with fixed (large) amplitude, and as a function of quench amplitude at fixed (short) duration. The most important feature shown is the appearance of simple power law scaling regimes. Combining the results of the two plots, one finds that in the limit of abrupt quenches (where the quench width is much smaller than all other dimensionful scales)
\begin{equation}
\langle T_{tt}\rangle_\mathrm{FINAL} \sim \left(\frac{\tilde{\delta}}{\tilde{\tau}}\right)^2.
\end{equation}

In fact this simple scaling relation was already anticipated on very general grounds, and is a special case of the universal scaling formula in \cite{Buchel:2013gba} for a
dimension three scalar operator.
That this scaling should be universal readily follows from the fact that for very fast quenches, the perturbation does not have time to propagate far from the boundary before the quench concludes. Since the near boundary region of many holographically relevant spacetimes is asymptotically AdS, abrupt quenches are indifferent to the IR features of the geometry distinguishing different holographic spacetimes.

\section{Discussion}
\label{sec:discuss}
One of the most important lessons to be extracted from our results is the domination of the thermalization time by the linear regime, even in the presence of a confinement scale in the holographic gauge theory. Indeed, in every quench that we have managed to evolve in this system, the thermalization time appears to be controlled entirely by the lowest lying quasi normal mode of the final state black brane.  It is conceivable that this is due in large part to the initial state that we have chosen to perturb. As a consequence of the previously discussed IR features of very small black holes in this model, our numerical stability suffers as the size of the initial state black brane shrinks. On the other hand, it is sensible to expect that the confinement scale becomes important to the characteristics of the dynamical response when it is the dominant energy scale in the problem. Currently, the smallest black hole that we can reliably perturb has an energy density comparable to the confinement scale: $\langle T_{tt} \rangle /f_0\,^4 \sim O(1)$. Thus, it is perhaps not entirely surprising that the dynamical response of the system does not appear to pass through any regime controlled by this energy scale in our computations.

Ultimately, it would be of great interest to perturb initial states with $\langle T_{tt} \rangle /f_0\,^4 \ll1$, or better yet the zero temperature solution itself. It is within this setting that one expects the response to the quench to be most sensitive to the diverging scalar potential in the bulk, and thus to the confinement scale of the dual boundary theory. Understanding to what extent this scalar potential is similar to the hardwall in its ability to induce scattering solutions which never collapse in the bulk is perhaps the most obvious avenue to pursue. Should such scattering solutions exist, identifying the boundary in the space of quench parameters that separates perturbations which lead to horizon formation from those which do not would permit a scaling analysis analogous to that performed in the well known case of Choptuik phenomena \cite{Choptuik:1992jv}.

Last, it may prove interesting to investigate the features of other probes sensitive to the quench dynamics in this holographic theory. Some obvious contenders are various non-local probes such as Wilson loops, entanglement entropy, and two point functions of large dimension operators. These operators all share the feature that their holographic computation involves the study of worldlines or worldsheets that sag into the bulk spacetime. As the separation of boundary operators increases, these surfaces generically droop lower in the radial (holographic) direction. In this way, these probes may provide insights into the equilibration of the dual field theory at different length scales, potentially offering a more detailed understanding of the process.

\acknowledgement

This work was supported in part by European Union's Seventh Framework Programme
under grant agreements (FP7-REGPOT-2012-2013-1) no 316165, the EU program ``Thales" MIS 375734
 and was also cofinanced by the European Union (European Social Fund, ESF) and Greek national funds through
the Operational Program ``Education and Lifelong Learning" of the National Strategic
Reference Framework (NSRF) under ``Funding of proposals that have received
a positive evaluation in the 3rd and 4th Call of ERC Grant Schemes".
The work of T.I.~was supported in part by the Department of Energy, DOE award No.~DE-SC0008132.
The work of C.R.~was supported in part by the European Research Council under the European Union's Seventh Framework Programme (FP7/2007-2013), ERC Grant agreement ADG 339140.

\bibliography{bibicnfp2015}

\end{document}